\documentclass[letter,superscriptaddress,twocolumn,prl,showkeys,showpacs]{revtex4-1}

\usepackage[utf8]{inputenc}
\usepackage{amsmath}
\usepackage{amssymb}
\usepackage{amsfonts}     
\usepackage{graphicx}   
\usepackage{verbatim}  
\usepackage{color}     
\usepackage{subfigure}  
\usepackage{hyperref}   
\usepackage{float}
\usepackage{natbib}
\usepackage{gensymb}
\usepackage{color}
\usepackage{epstopdf}
\usepackage{multirow}
\usepackage{xfrac}

\begin{document}

\title{Semimetal-to-semiconductor transition and charge-density-wave melting in $1T$-TiSe$_{2-x}$S$_x$ single crystals}
\author{M.-L. Mottas}
\altaffiliation{Corresponding author.\\ marie-laure.mottas@unifr.ch}
\affiliation{D{\'e}partement de Physique and Fribourg Center for Nanomaterials, Universit{\'e} de Fribourg, CH-1700 Fribourg, Switzerland}

\author{T. Jaouen}
\altaffiliation{Corresponding author.\\ thomas.jaouen@unifr.ch}
\affiliation{D{\'e}partement de Physique and Fribourg Center for Nanomaterials, Universit{\'e} de Fribourg, CH-1700 Fribourg, Switzerland}

\author{B. Hildebrand}
\affiliation{D{\'e}partement de Physique and Fribourg Center for Nanomaterials, Universit{\'e} de Fribourg, CH-1700 Fribourg, Switzerland}

\author{M. Rumo}
\affiliation{D{\'e}partement de Physique and Fribourg Center for Nanomaterials, Universit{\'e} de Fribourg, CH-1700 Fribourg, Switzerland}

\author{F. Vanini}
\affiliation{D{\'e}partement de Physique and Fribourg Center for Nanomaterials, Universit{\'e} de Fribourg, CH-1700 Fribourg, Switzerland}

\author{E. Razzoli}
\affiliation{Quantum Matter Institute, University of British Columbia, Vancouver, BC, Canada V6T 1Z4}
\affiliation{Departement of Physics and Astronomy, University of British Columbia, Vancouver, BC, Canada V6T 1Z1}

\author{E. Giannini}
\affiliation{Department of Quantum Matter Physics, University of Geneva, 24 Quai Ernest-Ansermet, 1211 Geneva 4, Switzerland}

\author{C. Barreteau}
\affiliation{Department of Quantum Matter Physics, University of Geneva, 24 Quai Ernest-Ansermet, 1211 Geneva 4, Switzerland}

\author{D. R. Bowler}
\affiliation{London Centre for Nanotechnology and Department of Physics and Astronomy, University College London, London WC1E 6BT, UK}

\author{C. Monney}
\affiliation{D{\'e}partement de Physique and Fribourg Center for Nanomaterials, Universit{\'e} de Fribourg, CH-1700 Fribourg, Switzerland}

\author{H. Beck}
\affiliation{D{\'e}partement de Physique and Fribourg Center for Nanomaterials, Universit{\'e} de Fribourg, CH-1700 Fribourg, Switzerland}

\author{P. Aebi}
\affiliation{D{\'e}partement de Physique and Fribourg Center for Nanomaterials, Universit{\'e} de Fribourg, CH-1700 Fribourg, Switzerland}

\begin{abstract}

The transition metal dichalcogenide $1T$-TiSe$_2$ is a quasi-two-dimensional layered material with a phase transition towards a commensurate charge density wave (CDW) at a critical temperature T$_{c}\approx 200$K. The relationship between the origin of the CDW instability and the semimetallic or semiconducting character of the normal state, i.e., with the non-reconstructed Fermi surface topology, remains elusive. By combining angle-resolved photoemission spectroscopy (ARPES), scanning tunneling microscopy (STM), and density functional theory (DFT) calculations, we investigate $1T$-TiSe$_{2-x}$S$_x$ single crystals. Using STM, we first show that the long-range phase coherent CDW state is stable against S substitutions with concentrations at least up to $x=0.34$. The ARPES measurements then reveal a slow but continuous decrease of the overlap between the electron and hole ($e$-$h$) bands of the semimetallic normal-state well reproduced by DFT and related to slight reductions of both the CDW order parameter and $T_c$. Our DFT calculations further predict a semimetal-to-semiconductor transition of the normal state at a higher critical S concentration of $x_c$=0.9 $\pm$0.1, that coincides with a melted CDW state in TiSeS as measured with STM. Finally, we rationalize the $x$-dependence of the $e$-$h$ band overlap in terms of isovalent substitution-induced competing chemical pressure and charge localization effects. Our study highlights the key role of the $e$-$h$ band overlap for the CDW instability.

\end{abstract}
\date{\today}
\maketitle

\section {I. Introduction}
Quasi-two-dimensional transition metal dichalcogenides (TMDCs) are layered compounds exhibiting a wide variety of interesting electronic properties \cite{Wang2012a}, often undergoing charge density wave (CDW) transitions \cite{Rossnagel2011}, or/and hosting low-temperature superconductivity \cite{Castro_2001} upon electronic doping or pressure (mechanical or chemical as induced by isovalent substitution) \cite{Wang2015}. The TMDC $1T$-TiSe$_2$ is one of these materials which has been studied for decades. It undergoes a $2\times 2 \times 2$ charge density modulation with a weak periodic lattice distorsion (PLD) occuring around $T_{c}\approx 200$ K \cite{DiSalvo1976, Hildebrand2018a}. It is superconducting upon electrical gating \cite{Li2015}, Cu intercalation \cite{Morosan2006} or under pressure \cite{Kusmartseva2009}. Increasing the crystal growth temperature further induces Ti self-doping which strongly perturbs the CDW phase coherence \cite{Hildebrand2016,Hildebrand2017} and drastically decreases the anomalous peak in the temperature-dependent resisitivity curves \cite{DiSalvo1976}.

In contrast to $1T$-TiSe$_2$, no phase transition has been reported in $1T$-TiS$_2$ \cite{Chen1980,Isomaki_1981, Wang_1991}. It is therefore expected to observe a gradual suppression of the CDW in the ternary compound $1T$-TiSe$_{2-x}$S$_x$ for increasing $x$. Studies of $1T$-TiSe$_{2-x}$S$_x$ have already been carried out, mainly based on temperature-dependent resistivity measurements \cite{DiSalvo1976,Lopez-Castillo1987,Miyahara1996,May2011}, but also with Raman scattering \cite{Freund1984}, tunneling spectroscopy \cite{Miyahara1996} and angle-resolved photoemission spectroscopy (ARPES) \cite{May2011}. All these investigations concluded that the S substitution monotonically lowers $T_c$ until a critical concentration $x_c$ close to 1.

Here, we re-investigate the CDW melting in $1T$-TiSe$_{2-x}$S$_x$ single crystals by combining ARPES, scanning tunneling microscopy (STM), and density functional theory (DFT) calculations, thereby giving a direct view of both real-space and electronic band structures and characterizing the crystals extremely well with respect to native defects and S concentrations. We demonstrate that in contrast to the electron-donors Ti and Cu intercalants that shift the 1$T$-TiSe$_2$ chemical potential \cite{Jaouen2018, Qian2007}, isovalent S-substitutions reduce the electron-hole ($e$-$h$) band overlap in the normal state and only slightly affect the long-range phase coherent CDW state as long as the $1T$-TiSe$_2$ normal state remains semimetallic. Our DFT analysis reveals that whereas the isovalent S substitution induces an increase of the overlap of the Ti 3$d$ electron and the Se 4$p$ hole bands by positive chemical pressure, charge localization effects as introduced by the more localized S 3$p$ orbitals leads to a reduced $p$-$d$ hybridization. The charge localization effect dominates the structural counterpart of the S substitution and drives the $e$-$h$ band gap opening. The CDW is experimentally found to be melted at $x$ $\sim$1 in good agreement with the DFT-predicted semimetal-to-semiconductor transition, therefore demonstrating that the Fermi surface has to host electron and hole pockets for driving the CDW instability. 

\section {II. Experimental details}
The $1T$-TiSe$_{2-x}$S$_x$ single crystals were grown by iodine vapor transport and contain less than  $0.2\%$ of intercalated Ti as measured by STM. Constant current STM images were recorded at $4.5$ K with a Omicron low-temperature (LT)-STM in constant current mode by applying a bias voltage V$_{bias}$ to the sample. The base pressure was better than $5\times10^{-11}$mbar. The temperature-dependent ARPES study was carried out using a Scienta DA$30$ photoelectron analyzer and monochromatized He-I radiation as exciting source ($h\nu$=$21.22$ eV, SPECS UVLS with TMM 304 monochromator). The total energy resolution was of the order of $12$ meV and the sample temperatures were deduced from fits of the Fermi edge spectra on the Cu sample holder. 

\section {III. Computational method}
DFT model calculations were performed using the plane-wave pseudopotential code VASP \cite{Kresse1993, Kresse1996}, version 5.3.3. Projector augmented waves \cite{Kresse1999} were used with the Perdew-Burke-Ernzerhof (PBE) \cite{Perdew1996} exchange correlation functional. The cell size of our model was 28.035 \AA~$\times$ 28.035 \AA. The 1$T$-TiSe$_2$ surface was modeled with two layers and the bottom Se layer fixed. A Monkhorst-Pack mesh with 2$\times$2$\times$1 $k$ points was used to sample the Brillouin zone of the cell. The parameters gave an energy difference convergence of better than 0.01 eV. During structural relaxations, a tolerance of 0.03 eV/\AA~ was applied. STM images were generated using the Tersoff-Hamann \cite{Tersoff1998} approach in which the current $I(V)$ measured in STM is proportional to the integrated local density of states (LDOS) of the surface using the BSKAN code \cite{Hofer2003}.

In order to compute the electronic band structures measured in ARPES, DFT calculations were done using the WIEN2K package \cite{wien2k} with the modified Becke-Johnson (mBJ) exchange-correlation potential \cite{Tran_2009}. This functionnal contains a tuning parameter $c_{mBJ}$ [see Ref. \cite{Koller_2012}]which has ben set to 1.34 to give the best agreement to the measured room-temperature (RT) $1T$-TiSe$_2$ pristine band structure. We used a $2\times 2 \times 2 $ superstructure of $8$ unit cells of $1T$-TiSe$_2$ with in-plane and out-of-plane experimental lattice parameters $a$= 3.54 \AA\ and $c$= 6.01 \AA\ \cite{DiSalvo1976}, respectively. By replacing $1, 2, 3$ and $4$ Se atoms with S, we obtained S concentrations of $x$ = 0.125, 0.25, 0.375, and 0.5, respectively. The associated unit cell parameters were set to $a$= 3.53, 3.52, 3.51, and 3.50 \AA\; and $c$= 5.99, 5.97, 5.95, and 5.93 \AA\ , respectively \cite{Miyahara1996}. The calculated band structures were unfolded using the FOLD2BLOCH package \cite{Rubel_2014}.

\begin{figure}[t]
\includegraphics[width=0.5\textwidth]{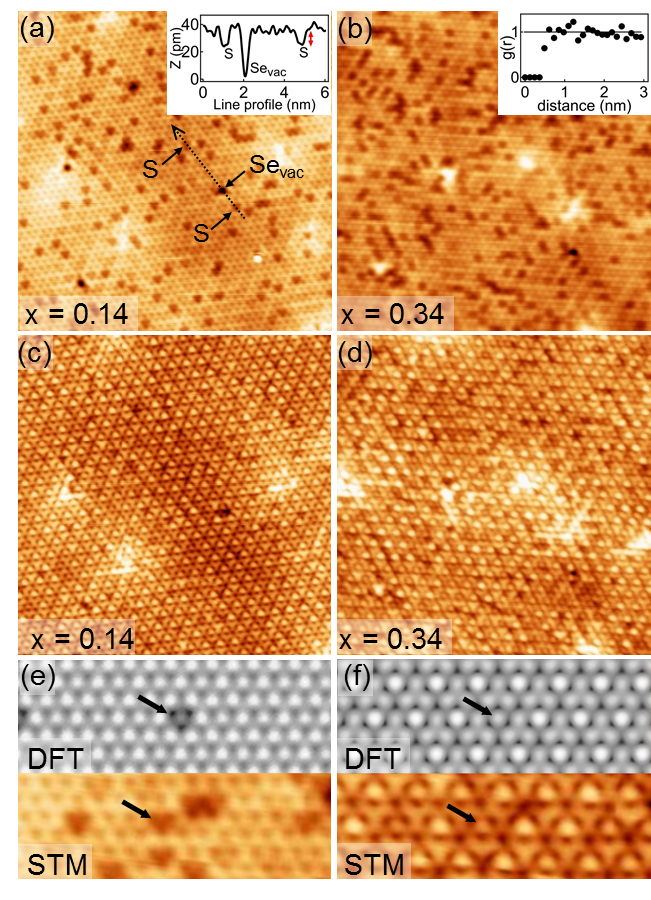}
\caption{\label{fig1} (Color online) $15\times 15$ nm$^2$ constant current (I$=0.15$ nA) STM images of $1T$-TiSe$_{2-x}S_x$ recorded at $T=4.5$ K. Applied bias voltage is V$_{bias}=+0.6$ V for (a) and (b). Insets show a line profile from image (a) and the normalized radial distribution function g(r) of sulfur atoms from image (b). (c), (d) Same surface regions as (a), (b) with V$_{bias}=+0.15$ V in order to highlight the $2 \times 2$ charge density modulation. (e), (f) Comparison of DFT-simulated and experimental $5\times3$ nm$^2$ STM images with sulfur substitutions, for $V_{bias}=+0.6$ V (e) and $V_{bias}=$+0.1 V (DFT) and +0.15 V (STM) (f). The arrows on (e) and (f) show the location of one S substitution.
} 
\end{figure}

\begin{figure*}[t]
\includegraphics[width=1\textwidth]{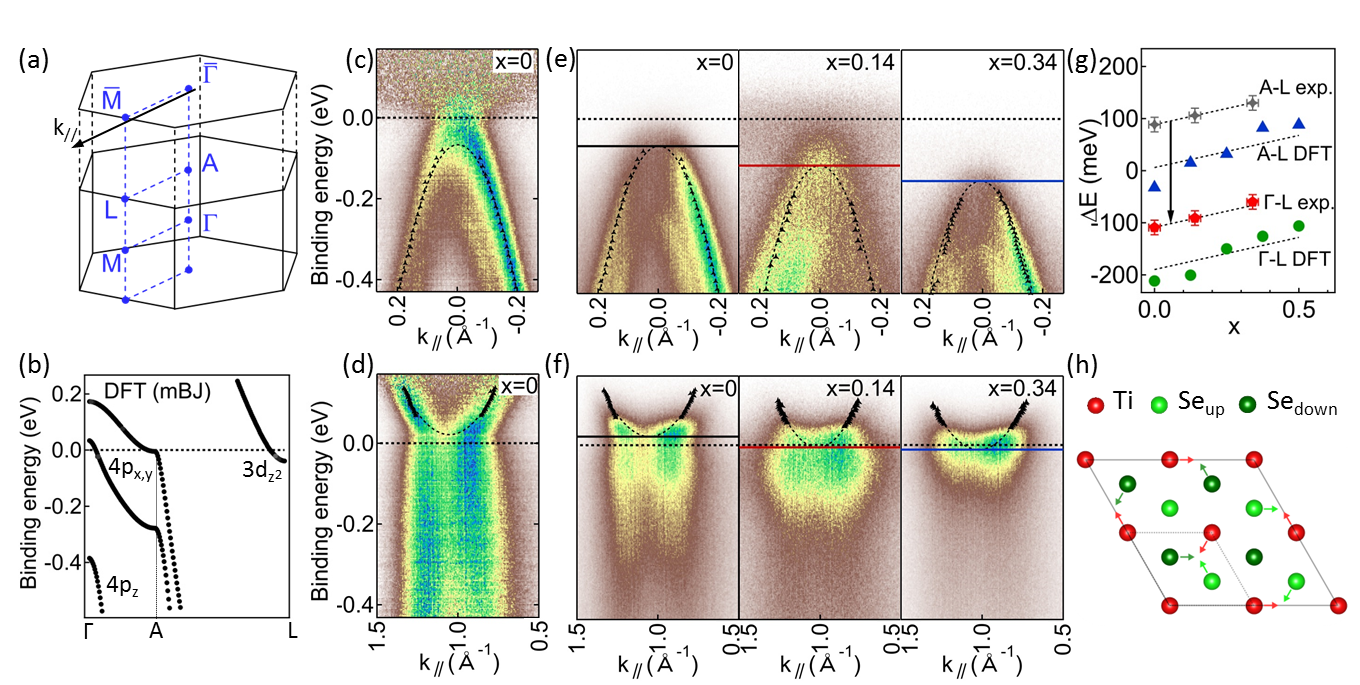}
\caption{\label{fig2} (Color online) (a) $1T$-TiSe$_{2}$ bulk and surface Brillouin zones. (b) DFT electronic band structure along the $\Gamma$-A-L path in the normal phase. (c), (d) Room temperature (RT) MDC-normalized ARPES spectra of pristine $1T$-TiSe$_{2}$ at the $\overline{\Gamma}$ and $\overline{M}$ points of the surface BZ, respectively. (e)  ARPES spectra of $1T$-TiSe$_{2-x}$S$_x$ crystals for $x=0, 0.14$ and $0.34$ (from left to right) displaying the valence band at $\overline{\Gamma}$. (f) RT ARPES spectra of the conduction band recorded at $\overline{M}$ as a function of $x$ (from left to right). (g) Comparisons of the energy difference between the valence band at $\Gamma/A$ and conduction band at $L$ as a function of $x$ from DFT calculations and experiment. Negative and positive values respectively refer to $e$-$h$ band overlap and band gap. (h) Top-view of the $1T$-TiSe$_{2}$ structure and lattice deformation associated with the CDW.} 
\end{figure*}

\section {IV. Results and Discussion}

Figure \ref{fig1}(a) and (b) show STM images recorded at +0.6 V and $T$=4.5 K. At this V$_{bias}$, the $2\times 2$ electronic modulation is not resolved because the CDW phase transition mainly affects the $1T$-TiSe$_{2}$ density of states close to the Fermi level. The S atoms are distinguishable as depletions at the location of Se atoms of the topmost $1T$-TiSe$_{2}$ layer, with S concentrations $x=0.14\pm0.02$ for Fig. \ref{fig1}(a) and $x=0.34\pm0.01$ for Fig. \ref{fig1}(b), determined with statistics made on several similar images. Well-known native defects are also present in negligible quantities \cite{Hildebrand2014}, including iodine and oxygen substitutions, as well as selenium vacancies which appear as depletions too. The distinction between S substitutions and Se vacancies is clear, as shown by the line profile in the inset of Fig \ref{fig1}(a). Regardless of the applied V$_{bias}$, S substitutions always appear as depletions with a $z$-value of about $14$ pm lower than the Se value, which corresponds to the difference of ionic radius sizes between the ions S$^{2-}$ and Se$^{2-}$ [see the red arrow in the inset of Fig \ref{fig1}(a)]. The electronegativity difference between S and Se being indeed very small, the S atom fingerprint is essentially topographic. The average radial distribution function $g(r)$ \cite{chaikin_lubensky_1995}, calculated for many different central S atoms, is displayed in the inset of Fig \ref{fig1}(b) and shows that S atoms are randomly distributed.

Figure \ref{fig1}(c) and (d) display STM images of the same zones as Fig \ref{fig1}(a) and (b) with V$_{bias}=+0.15$ V in order to highlight the commensurate CDW charge modulation. Whereas low concentrations of intercalated-Ti atoms significantly affect the long-range phase coherence of the CDW \cite{Hildebrand2017}, no phase shift was observed on these two S-substituted samples at low temperature. The phase coherence of the charge modulation is thus totally insensitive to S substitutions with concentrations at least up to $x=0.34$. Figure \ref{fig1}(e) and (f) compare DFT-simulated STM images with V$_{bias}=+0.6$ V (e) and V$_{bias}=+0.1$ V (f) and measured STM images with V$_{bias}=+0.6$ V (e) and V$_{bias}=+0.15$ V (f). Fig \ref{fig1}(e) confirms the S substitution identification at V$_{bias}=+0.6$ V. Also, even if at lower V$_{bias}$ the CDW modulation tends to hide the topographic depletions associated with S substitutions of non-displaced Se atoms of the PLD \cite{Hildebrand2017}, it is still possible to identify them as weakened CDW maxima [see black arrows on Fig \ref{fig1}(f)].

\begin{figure*}[t]
\includegraphics[width=1\textwidth]{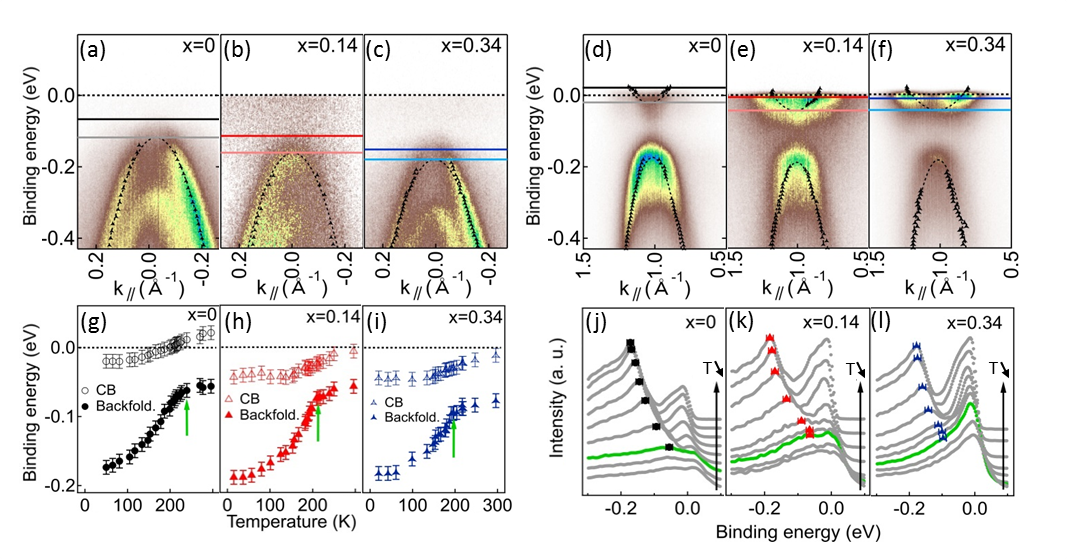}
\caption{\label{fig3} (Color online) (a)-(c) ARPES spectra of $1T$-TiSe$_{2-x}$S$_x$ for $x=0, 0.14$ and $0.34$ recorded at low temperature ($T=46\pm5$ K) and displaying the valence band at $\bar{\mathrm{\Gamma}}$. (d)-(f) Low temperature ARPES spectra measured at $\bar{M}$, where the backfolded band is well visible. The two lines on each spectrum (a)-(f) indicate the determined valence and conduction band extrema at room (upper lines) and low (lower lines) temperatures. (g)-(i) Evolution of the conduction band minimum and the backfolded weight maximum as a function of the temperature for the three crystals. (j)-(l) EDCs taken on spectra at $\bar{M}$ as a function of temperature, for $x=0, 0.14$ and $0.34$. The markers indicate the position of the backfolded spectral weight maximum and the green EDCs correspond to the maximum temperature at which it can still be identified (the corresponding extracted positions on (g)-(i) are indicated by the green arrows).} 
\end{figure*}

\begin{figure}[h]
\includegraphics[width=0.5\textwidth]{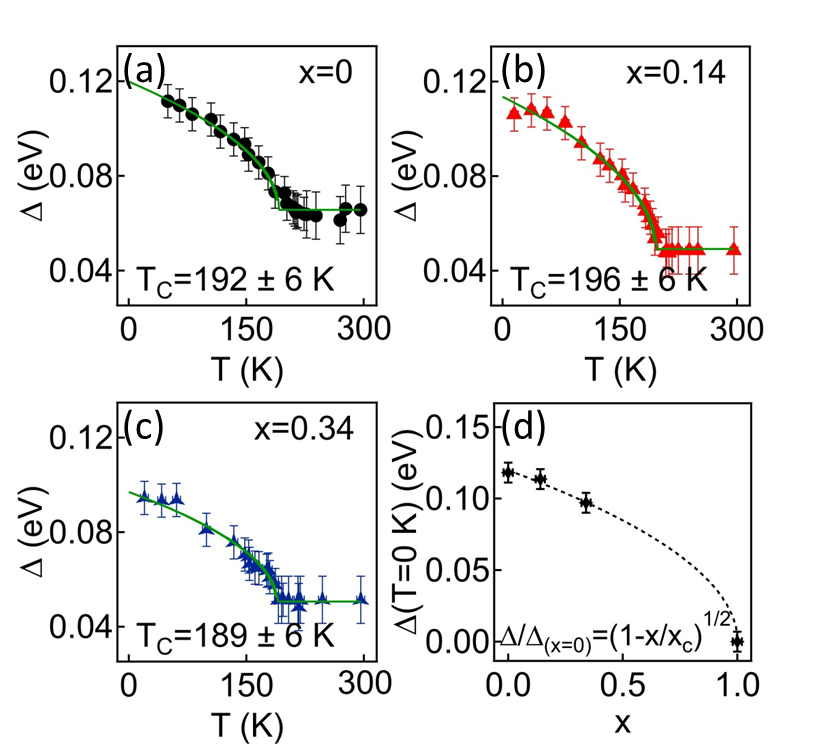}
\caption{\label{fig4} (Color online) (a)-(c) $T$-dependence of the CDW order parameter $\Delta$ deduced from ARPES for $x=0, 0.14$ and $0.34$, respectively. For each sample, $\Delta(T)$ is fitted with a mean-field-like power-law. (d) $x$-dependence of $\Delta(0)$. The black dashed line is a mean-field-like fit of $\Delta$ vs $T$ given by $\Delta/\Delta_{x=0}=(1-x/x_c)^{1/2}$ using only the three $\Delta_{x=0}$ values extracted from (a), (b) and (c) and giving a critical S concentration $x_c$=$1.0\pm0.3$.}  
\end{figure}

Let us now focus on the $x$-dependence of the 1$T$-TiSe$_2$ band structure. The $1T$-TiSe$_2$ low-energy electronic states consist of a Se 4$p$ hole-pocket and Ti 3$d$ electron-pockets at the $\mathrm{\Gamma}$ and three-equivalent L points of the three-dimensional (3D) Brillouin zone (BZ) [see Fig. \ref{fig2}(a)]. The DFT-calculated hole pocket is composed of two weakly dispersing $p_{x,y}$ bands along $\mathrm{\Gamma}$-$A$ and one strongly dispersing band of mainly $p_z$ character at higher binding energy  \cite{Zuzana2015}, whereas the Ti 3$d$ electron-pocket is of mainly $d_{z^2}$ character [see Fig. \ref{fig2}(b)]. 

Fig. \ref{fig2}(c) and (d) show RT ARPES spectra at the $\bar{\mathrm{\Gamma}}$ and $\bar{M}$ points of the hexagonal surface BZ [see Fig. \ref{fig2} (a)] of pristine 1$T$-TiSe$_2$ normalized according to momentum distribution curves (MDCs) for a better visualization of the electron and hole band dispersions. These (black dashed curves) are obtained by Lorentzian-fitting of MDCs [black markers on (c) and (d)] throughout our ARPES study. Since He-I ARPES is mainly probing the $A-L$ plane of the 3D BZ (in a free-electron final-state picture) \cite{Pillo2000}, only the Se 4$p_{x,y}$ bands are seen in our $\bar{\mathrm{\Gamma}}$ ARPES spectra in Fig. \ref{fig2}(e). Whereas the binding energy of the Se 4$p$ band (indicated by horizontal lines) strongly increases with $x$, the bottom of the Ti 3$d$ electron band at $\bar{M}$ is only slightly shifted [Fig \ref{fig2}(f)]. Thus, in contrast to the electron-donors Ti and Cu intercalants that shift the 1$T$-TiSe$_2$ chemical potential \cite{Jaouen2018, Qian2007}, isovalent S-substitutions change the RT $e$-$h$ band overlap. 

More precisely, the experimental $\bar{\mathrm{\Gamma}}$-$\bar{M}$ band gap is found to linearly increase with $x$ [$A$-$L$ exp. on Fig. \ref{fig2}(g)] with a slope in very good agreement with those of both the DFT-calculated $A$-$L$ band gap and $\Gamma$-$L$ band overlap variations [blue and green markers on Fig. \ref{fig2}(g), respectively]. This indicates first, that the DFT-calculated $k_z$ dispersion along $\Gamma$-$A$ remains rather insensitive to S substitutions. Second, to obtain a \textit{semi}-empirical S-concentration dependence of the $\Gamma$-$L$ band overlap, we consider that DFT well reproduces the experimental $k_z$ dispersion along $\Gamma$-$A$ and we shift the experimental $\bar{\mathrm{\Gamma}}$-$\bar{M}$ values by the energy difference in DFT between the top of the hole pocket at $\Gamma$ and $A$ [red markers and arrow on Fig. \ref{fig2}(g)]. 

At low temperature, in the CDW phase, the $\mathrm{\Gamma}$ and $L$ points are connected by the new reciprocal lattice vectors corresponding to the doubling of the lattice periodicity and large band renormalizations at high-symmetry points of the BZ as well as a large transfer of spectral weight to backfolded bands appear in ARPES. The $p-d$ hybridization is strongly orbital and $k_z$-dependent \cite{Watson2018} and mainly leads to the renormalization of the Se $4p$ bands at $\mathrm{\Gamma}$ and Ti $3d$ bands at $L$ that derive from the atoms involved in the Ti-Se bond shortening of the PLD [Fig. \ref{fig2} (h)]. 

ARPES spectra at $\bar{\mathrm{\Gamma}}$ in the CDW phase are displayed on Fig \ref{fig3}(a), (b) and (c) for $x$=0, 0.14 and 0.34 respectively and show Se 4$p$ bands that have shifted to higher binding energy upon temperature lowering [see the two horizontal lines Fig. \ref{fig3}(a)-(c)]. These shifts are mainly associated with the temperature-induced chemical-potential shift $\Delta \mu(T)$ arising from the effect of the Fermi-Dirac cutoff on the overlapped hole and electron bands with nonequal band masses \cite{Claude2009}. Focusing on the $x$-dependent evolutions of both the Se 4$p$ backfolded band and the Ti 3$d$ electron pocket at $\bar{M}$ in the CDW state [Fig. \ref{fig3}(d)-(f)], we first recognize the characteristic backfolded Se 4$p$ bands coming from $\mathrm{\Gamma}$ for all three samples with yet a diminished spectral weight for $x=0.34$. Then, we also observe slight shifts of the electron pocket minimum with respect to the RT ARPES spectra. The electron pocket minimum is not influenced by the CDW transition since the lowest Ti 3$d$ conduction band in the CDW phase derives from Ti atoms that do not move with the PLD \cite{Hellgren2017} [Fig. \ref{fig2}(h)]. In fact, the observed energy shift of the electron pocket minimum also fully relates to $\Delta \mu(T)$. It therefore depends on the RT $e$-$h$ band overlap as exemplified by the decreasing $\Delta\mu$ values of $\sim$41, $\sim$38, and $\sim$33 meV for $x$=0, 0.14 and 0.34, respectively obtained for the energy shift of the Ti 3$d$ bands from room to low temperature [as indicated by the two horizontal lines Fig. \ref{fig3}(d)-(f)].

The detailed $T$-evolution of the Ti 3$d$ and backfolded Se 4$p$ bands extrema of our three $1T$-TiSe$_{2-x}$S$_x$ crystals are indicated Fig. \ref{fig3}(g)-(i). They have been obtained from parabolic fits [dashed lines Fig. \ref{fig3}(d)-(f)] of the MDC maxima determined on the ARPES data [black symbols Fig. \ref{fig3}(d)-(f)]. The effective masses have been kept fixed at the RT values for all temperatures, and the initial band extrema positions adjusted to the maxima of the EDCs taken exactly at $\bar{M}$ as shown in Fig. \ref{fig3}(j)-(l). We indeed see the previously discussed $\Delta \mu(T)$ on the Ti 3$d$ bands, and the downward shift of the Se 4$p$ backfolded bands accompanied by continuous increases of their spectral weight. The green curves are EDCs obtained at the maximum temperature at which the backfolding spectral weight can be straightforwardly identified, i.e., approximatively at the critical CDW temperatures [the corresponding Se $4p$ backfolded band positions are indicated by the green arrows on (g)-(i)].

\begin{figure}[b]
\includegraphics[width=0.5\textwidth]{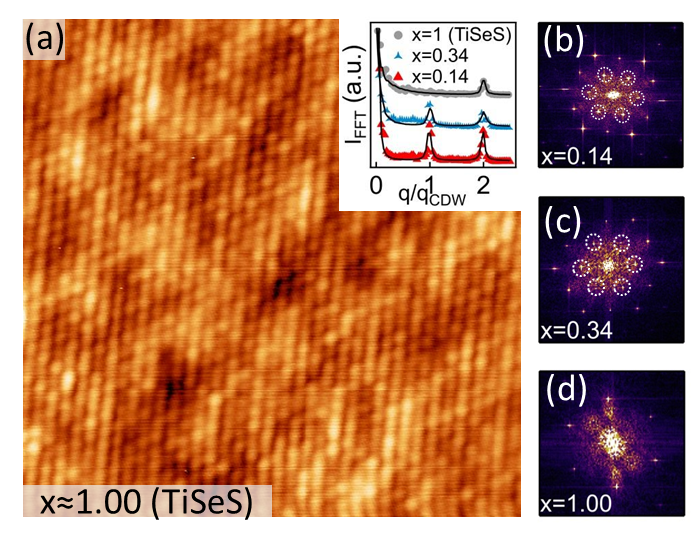}
\caption{\label{fig5} (Color online) (a) $11 \times 11$ nm$^2$ constant current (I$=0.10$ nA) STM image of $1T$-TiSe$_{2-x}$S$_x$ with $x\approx1$ recorded at $T=4.5$ K with applied bias voltage $V_{bias}=+0.05$ V. The inset shows line profiles of Fast-Fourier-Transform (FFT) spots taken on the FFT-amplitude plots (b)-(d) obtained from STM images of the $1$ $1T$-TiSe$_{2-x}$S$_x$ crystals in Fig. \ref{fig1}(c), (d) for $x=0.14, 0.34$ and in Fig. \ref{fig5}(a) for $x\approx1$, respectively. The white circles show the extra spots originating from the CDW. For $x\approx1$, the extra spots coming from the $2\times2$ charge modulation are not present.}  
\end{figure}

Within a BCS-like approach, the shift of the backfolded band at $L$ (corrected by $\Delta \mu(T)$) relates to the order parameter $\Delta$ describing the coupling strength between the valence band at $\mathrm{\Gamma}$ and three symmetry equivalent conduction bands at $L$ \cite{Claude2009, Monney2009_bis}. Figures \ref{fig4}(a), (b) and (c) show the $T$-dependence of $\Delta$ obtained from ARPES using our semi-empirical values of the $e$-$h$ band overlap. Mean-field-like fits to the $\Delta$ values with $\Delta(T)=\Delta_0(1-T/T_c)^{1/2}+\Delta(T_c)$ gives rather similar $T_c$ values of $192\pm6$, $196\pm6$ K and $189\pm6$ K for $x$=0, 0.14$\pm0.02$ and 0.34$\pm0.01$, respectively, again demonstrating that the CDW phase transition is stable against S substitutions in that range of concentrations. Nevertheless, our ARPES measurements also reveal a slow but continuous decrease of $\Delta(0)$ [ $0.124\pm0.007, 0.118\pm0.007$ and $0.101\pm0.007$ eV for $x=0, 0.14\pm0.02$ and $0.34\pm0.02$, respectively]. Focusing on the $x$-dependence of $\Delta(0)$, Fig. \ref{fig4}(d), and considering a mean-field like scaling behaviour, we anticipate a complete suppression of the CDW phase transition at a critical sulfur concentration $x_c$ of $1.0\pm0.3$, which is in good agreement with the critical value given in Ref. \cite{DiSalvo1976}.  

As a confirmation, we show, on Fig. \ref{fig5}, LT-STM results obtained on a S-substituted $1T$-TiSe$_2$ crystal with $x\approx1$ and negligible intercalated-Ti concentration. With $V_{bias}=+0.05$ V, the CDW modulation at 4.5 K is not distinguishable. Comparing line profiles of the $(2\times2)$ and $(1\times1)$ Fast-Fourier-Transform (FFT) spots for $x=0.14, 0.34$ and $1$ [inset Fig. \ref{fig5} (a)], taken on Fig. \ref{fig5}(b), (c) and (d), respectively, we see that the peak at $q=q_{cdw}$ corresponding to the $2\times2$ charge modulation is well visible for $x=0.14$ and $x=0.34$ but does not exist for $x\approx1$, demonstrating that $\Delta(0)$ is close to zero at the value of $x_c$ deduced from ARPES data.

\begin{figure}[t]
\includegraphics[width=0.45\textwidth]{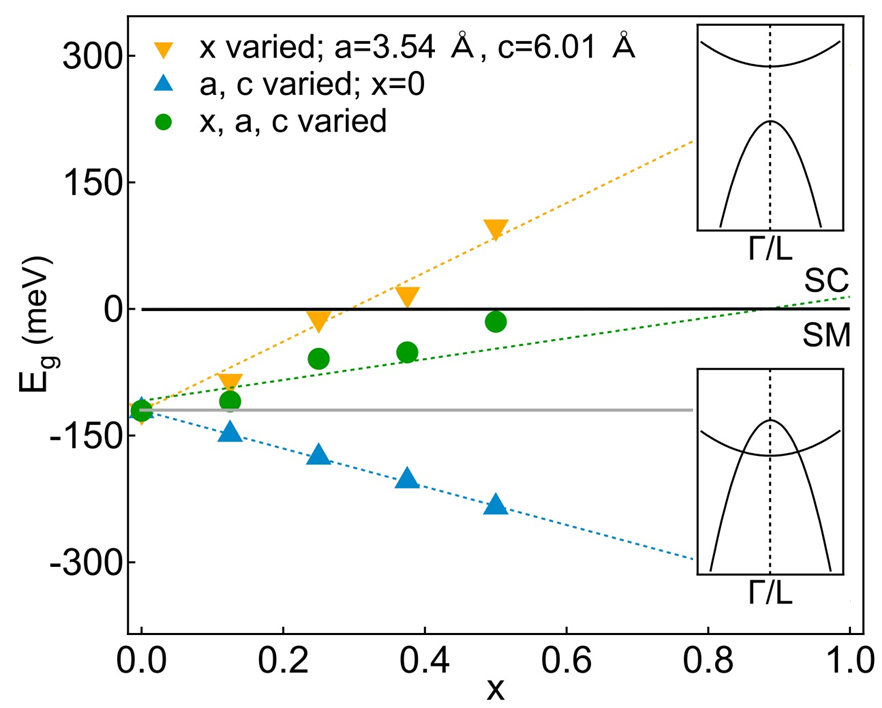}
\caption{\label{fig6} (Color online) Energy difference between the electron band minimum at $L$ and the hole band maximum at $\Gamma$ ($E_g$), as obtained by DFT for $x$ = 0, 0.125, 0.25, 0.375, and 0.5 and lattice parameters $a$ and $c$ fixed at their experimental values taken from \cite{Miyahara1996} (green circles). $E_g$ $<$0 and $E_g$ $>$0 respectively refer to semimetallic and semiconducting normal state. Are also shown the computed $E_g$ values obtained either by keeping fixed $a$ and $c$ to their pristine values while varying the S content (orange triangles) or from pristine $1T$-TiSe$_2$ by only changing $a$ and $c$ (blue triangles). The dashed-lines are linear fits to the $E_g$ values that have been all shifted to match our ARPES-extracted values.}  
\end{figure}

At this stage, we have shown that slight S substitutions induce a slow but continuous increase (decrease) of the $e$-$h$ band gap (overlap) of the semimetallic normal-state, well reproduced by DFT and that the CDW melts at a rather high $x_c$. We now show that the reductions of both the CDW order parameter $\Delta$ and $T_c$ are intimately related to the $e$-$h$ band gap opening. Figure \ref{fig6} displays the DFT-calculated energy difference between the electron band minimum at $L$ and the hole band maximum at $\Gamma$ ($E_g$), as a function of $x$, shifted to match our ARPES-extracted semi-empirical values [green markers, see also Fig. \ref{fig2}(g)]. Interestingly, extrapolating the linear dependence of $E_g$ with $x$ to higher substitutions, a semimetal-to-semiconductor transition ($E_g=$0) of the normal state is expected at a critical S concentration of 0.9 $\pm$0.1, a value not only very close to those proposed in previous temperature-dependent resistivity and Raman scattering measurements \cite{DiSalvo1976,Lopez-Castillo1987,Miyahara1996,May2011,Freund1984}, but especially coinciding with the concentration $x_c$ at which the CDW is melted. 

At last, our DFT analysis reveals the underlying mechanisms responsible for the $e$-$h$ band gap opening and the CDW melting. We first mimic chemical pressure effects introduced by isovalent S substitutions by calculating the electronic band structures of $1T$-TiSe$_2$ for couples of $a$ and $c$ lattice parameters corresponding to the experimental values of Ref. \cite{Miyahara1996}. As it can be seen in Fig.\ref{fig6} (blue markers), the overlap of the Ti 3$d$ electron and the Se 4$p$ hole bands linearly increases when the lattice contracts as expected from positive chemical pressure effects. On the other hand, keeping fixed the lattice parameters to their pristine values while varying only the S content, leads to computed $E_g$ that evolves almost linearly with $x$ towards a semiconducting band configuration (orange triangles, Fig.\ref{fig6}). Indeed, substituting the Se 4$p$ orbitals by the more localized S 3$p$ ones strengthens the ionic character of the ionocovalent transition-metal-chalcogen bonds, therefore acting as a charge localization effect and leading to a reduced $p$-$d$ hybridization. Overall, as seen in Fig. \ref{fig6}, charge localization dominates the chemical pressure effect and therefore drives the $e$-$h$ band gap opening, i.e. the CDW melting.

\section {V. Conclusion}

In summary, we have investigated the CDW melting in well characterized $1T$-TiSe$_{2-x}$S$_x$ single crystals by means of ARPES, STM, and DFT calculations. We have demonstrated that isovalent S-substitutions reduce the $e$-$h$ band overlap in the normal state and do not affect the long-range phase coherent CDW state as long as the $1T$-TiSe$_2$ normal state remains semimetallic. The CDW has been experimentally found to be melted at $x$ $\sim$1 in good agreement with the DFT-predicted semimetal-to-semiconductor transition. Our DFT analysis has revealed that whereas the isovalent S substitution induces an increase of the overlap of the Ti 3$d$ electron and the Se 4$p$ hole bands by positive chemical pressure, charge localization effects lead to a reduced $p$-$d$ hybridization and dominates the $e$-$h$ band gap opening. The mechanism of the CDW melting in $1T$-TiSe$_{2-x}$S$_x$ therefore indicates that only a semimetallic normal-state Fermi surface is unstable towards the $2\times 2 \times 2 $ CDW phase.
  
\begin{acknowledgments}
This project was supported by the Fonds National Suisse pour la Recherche Scientifique through Div. II. E. Razzoli acknowledges support from the Swiss National Science Foundation (SNSF) Grant No. P300P2-164649. Skillful technical assistance was provided by F. Bourqui, B. Hediger and O. Raetzo. 

\end{acknowledgments}

\bibliography{library1}
\end{document}